%% file: main.tex
\documentclass[letterpaper,twocolumn,10pt]{article}
\usepackage{usenix}
\usepackage[T1]{fontenc}
\usepackage[latin1]{inputenc}
\usepackage{here}
\usepackage{url}

\usepackage{booktabs}
\usepackage{multirow}

\usepackage{graphicx}
\graphicspath{{figs/}}
\usepackage{subfigure}

\usepackage{multirow}

\usepackage{pifont}
\newcommand{\cmark}{\ding{51}}%
\newcommand{\xmark}{\ding{56}}%

\author{
  {\rm Matthias W\"ahlisch}\\
  Freie Universit\"at Berlin
  \and
  {\rm Andr{\'e} Vorbach}\\
  Deutsche Telekom AG
  \and
  {\rm Christian Keil}\\
  DFN-CERT
  \and
  {\rm Jochen Sch\"onfelder}\\
  DFN-CERT
  \and
  {\rm Thomas C. Schmidt}\\
  HAW Hamburg
  \and
  {\rm Jochen H. Schiller}\\
  Freie Universit\"at Berlin
  \and
  {\{m.waehlisch,jochen.schiller\}@fu-berlin.de, Andre
  Vorbach@telekom.de}\\{\{keil,schoenfelder\}@dfn-cert.de,
  t.schmidt@ieee.org}
}

\title{Design, Implementation, and Operation of a Mobile Honeypot}

\begin{document}

\maketitle

\begin{abstract}
Mobile nodes, in particular smartphones are one of the most relevant
devices in the current Internet in terms of quantity and economic impact.
There is the common believe that those devices are of special interest for
attackers due to their limited resources and the serious data they store.
On the other hand, the mobile regime is a very lively network environment,
which misses the (limited) ground truth we have in commonly connected
Internet nodes. In this paper we argue for a simple long-term measurement
infrastructure that allows for (1) the analysis of unsolicited traffic to
and from mobile devices and (2) fair comparison with wired Internet access.
We introduce the design and implementation of a mobile honeypot, which is
deployed on standard hardware for more than 1.5 years. Two independent
groups developed the same concept for the system. We also present
preliminary measurement results.

\end{abstract}

\input{intro}
\input{related-work}

\input{honeypot}
\input{results}
\input{conclusion}

\input{acks}

\begin{small}
\bibliographystyle{IEEEtran}
\bibliography{/bib/own,/bib/rfcs,/bib/ids,/bib/theory,/bib/layer2,/bib/internet,/bib/transport,/bib/overlay,/bib/vcoip,/bib/ngi,/bib/security}
\end{small}

\end{document}

%% file: intro.tex
\section{Introduction} \label{sec:einleitung}

Scanning Internet hosts to initiate a denial of service, to find an exploit,
or to discover an unsecured remote access is typically the first step of an
attack towards Internet devices. In former times those attacks have been
reserved to traditional server systems \cite{apt-bhs-07}. Today, not only
desktops but also mobile devices (e.g., smartphones) offer intentionally
external services.

Mobile phones are particularly threaten by attacks. They are almost always
connected with the Internet. Their limited resources do not allow the
application of commonly used security mechanisms. In addition, many end
users disable security barriers, which have been introduced by vendors,
when they root or jailbreak their mobile \cite{p-bsij-13}. From this
perspective it is reasonably to assume that attackers specifically target
on mobile devices.  A plethora of research discussed network-based
vulnerability of mobiles and proposed solutions (e.g., \cite{gwz-sad-04}),
but up until now unsolicited remote accesses to mobiles have not been
studied in detail. In this paper, we argue that a measurement
infrastructure is required which aims to quantify and to analyse the amount
of remote attacks on mobiles.

A common technique to study attack behavior is the deployment of honeypot.
A honeypot is a trap for collecting data from unauthorized system
access---in this analysis via IP---initiated by remote parties. However,
the term ``mobile honeypot'' is not well-defined and there is only very
limited work on the design of a measurement system that allows for both,
the analysis of the mobile as well as non-mobile world.
In this paper we extend our previous work \cite{wtkss-fimh-12} and make the
following core contributions: 
\begin{enumerate}
\item We introduce the detailed design and
implementation concepts of a mobile honeypot. The principles of the system
have been developed independently by two groups. It has been running for
approximately 1.5 years and is part of the early warning system of one of
the largest ISPs in Europe. The honeypot system abstracts from unnecessary
mobile aspects, which allows us to deploy the same software base on
standard PCs that are connected to different types of Internet access.

\item We report on preliminary measurement results. This includes a summary
of our current observations of attack behaviour on smartphones, as well as
a statistical analysis of unsolicited traffic \cite{zc-wrdut-12}. The
traffic measurement presents data from November 2012, which we compare
with common wired Internet access and with our results from January 2012
\cite{wtkss-fimh-12}. Surprisingly, we do not find a significant amount of
attacks specific to mobiles, which indicates that adversaries operate
almost independently of the actually captured host. 

\end{enumerate}

The vulnerability of smartphones is based on multiple aspects. This
paper concentrates on remote attacks via the Internet. One might argue
that a mobile operator usually do not assign public IP addresses to mobiles
and that NAT techniques protect the systems against malicious access.  We
think the mobile environment needs an early and continuous analysis as well
as appropriate tools. There are still operators providing public IP
addresses. With an increased deployment of IPv6 the IP address assignment
policy will change as several application scenarios requiring direct
access without NAT traversal.

The remainder of this paper is structured as follows. In
Section~\ref{sec:related-work} we introduce background and discuss related
work in the context of mobile honeypots. We present the design space,
implementation, and deployment aspects of the mobile honeypot system in
Section~\ref{sec:mobilehoneypot}. Our measurement study is discussed in
Section~\ref{sec:ergebnisse}. We conclude with an outlook in
Section~\ref{sec:zusammenfassung}.


%% file: related-work.tex
\section{Background and Related Work} \label{sec:related-work}

\subsection{Trapping Attackers with a Honeypot}

In contrast to other security measures that ultimately try to keep the
attacker out of the system, honeypots are meant to be compromised. Their
value lies in luring the attacker into entering a system and collecting
information on how this is done.

A honeypot is typically classified as low interaction or high interaction
honeypot and client or server honeypot. A \emph{low interaction honeypot}
primarily collects information about the attacker and detects known
attacks. The limited level of interaction between attacker and target is
achieved by not providing fully functional services but only emulations
thereof with known exploits. On the other hand, a \emph{high interaction
honeypot} provides a fully functional system. They are used to reveal
current and new attacks that do not have to be catered for when setting up
the honeypot.  Since the high-interaction honeypot is a fully functional
system, it has to be closely monitored for successful attacks to prevent
the attacker from using the honeypot to target other systems on the
network.  \emph{Server honeypots} provide vulnerable services to malicious
clients.  Their focus is in protocol and service specific vulnerabilities.
A server honeypot does not offer any legitimate services, any connection by
a client can be treated as an attack. \emph{Client honeypots} take the role
of a vulnerable client trying to find malicious servers.

\subsection{Wireless versus Mobile Honeypots}

Physical and virtual honeypots \cite{ph-vhfbt-08} have been studied in
detail, however, there is only little work in the field of mobile-related
honeypots. Mobile honeypots have to be distinguished from \emph{wireless
honeypots} \cite{s-hwhma-07}, \cite{aeaa-whsa-09}, which focus on the
attacks on the wireless technology.  The term \emph{mobile honeypot} is
used here referring to honeypots that focus on attacks on mobile
devices.\footnote{Note that the term ``mobile honeypot'' is also used to
describe other scenarios. Balachander Krishnamurthy \cite{k-mmhtu-04} uses
it to describe prefixes of darknet address space that (1) are advertised to
upstream ASes, making the information mobile, and (2) change aperiodically,
moving the darknet in the address space.} They can either be mobile
themselves in running on the mobile device in which case they would usually
be low interaction honeypots used for deception and detection of known
attacks.  This also greatly reduces the possibility of the device itself
being compromised. On the other hand they can be dedicated devices up to
high interaction solutions set up to expose unknown attacks.  Mobile
honeypots in the sense of honeypots focussing on mobile devices are for
example developed by the Chinese Chapter of the Honeynet Project
\cite{honeynetproject}.  They are using prototype deployments of honeypots
for Bluetooth, WiFi, and MMS. TJ OConnor and Ben Sangster built honeyM
\cite{cs-hfivh-10}, a framework for virtualized mobile device client
honeypots, which emulates in particular wireless technologies.  Mulliner
\emph{et al.} \cite{mll-phcsh-11} propose HoneyDroid, a specific mobile
honeyot that exclusively runs on smartphones.  We argue that those
approaches complicate the measurement across different types of systems. In
addition, they are only required if the hardware characteristics are
relevant for the study.


%% file: honeypot.tex
\section{Mobile Honeypot System} \label{sec:mobilehoneypot}

Our primary goal is the design of a measurement system that captures
traffic characteristics of malicious behaviour on mobile devices and allows
for comparison with non-mobile environments. In addition to these
statistical observations, we are interested in the more detailed procedure of
potential attackers (e.g., which software do they infiltrate). A common
technique is the application of a honeypot. In this section, we discuss
appropriate levels of abstraction to cover the mobile environment without
losing comparability with non-mobile setups.
The mobile honeypot has been designed and implemented coincidently by two different
groups. Both groups approached completely independently at the same conclusion.


\paragraph{Attacker Model}

In this paper, we concentrate on a system that analyzes malicious access
via the Internet on smartphones. We argue for a typical attacker model. The
attacker tries to compromise the smartphone via unsolicited remote
connections \cite{gwz-sad-04}, or captures the mobile using malware and
initiates further denial of service attacks to other mobiles or non-mobile
hosts \cite{tlorj-cbmim-09}, \cite{ffchw-smmw-11}. In any case, such remote
attacks are bound to the network layer and moreover do not address
specifics of mobile hardware, but solely target at the system level. The
adversary actively tries to find vulnerable nodes and may use additional
information such as IP topology data or web server logs to differentiate
mobile and non-mobile networks.

\subsection{Design}

The term ``mobile honeypot'' is not well-defined. The general design space
is based on the following three questions:
(Q1) Is it necessary that the probe runs on a mobile device---if yes
  which device type (notebook versus smartphones versus \dots)?
(Q2) Is it necessary that the honeypot runs on a mobile operating system
  (PC emulation versus mobile device)?
(Q3) To which network is the mobile honeypot connected (DSL network
  versus UMTS network versus \dots)?

According to the attacker model, there is no need to operate the mobile
honeypot on real smartphones. This reduces complexity in building the
honeypot and simplifies long-term operation.

As underlying operating system we decided for Linux. This has two advantages: (1)
Most of the currently deployed smartphones use the Android OS. We conducted
fingerprinting tests using the well-known tools Nmap and Xprobe, which try to
guess the operating system. Both tools cannot distinguish Android from
current Linux versions. (2) Using Linux enables us to re-use existing
honeypot tools independently of the deployment in mobile or non-mobile
scenarios. This allows us the ensure comparison between different systems.

An important change is the adjustment of the virtual file system that is
presented to the attacker. It should reflect the directory structure of a
typical Android system.

To increase the \emph{attractiveness} for an adversary, we account for
``rooted'' (or ``jailbreaked'') devices. A rooted device grants additional
system access to the user. It allows post installation of additional
services such as HTTP or file sharing. A honeypot system that intends to
capture tools introduced by the attacker need to emulate a rooted
smartphone. Note, considering rooted devices provides the attacker with
\emph{supplementary} features and thus does not exclude off-the-shelf
mobiles.

Regarding the third question we argue that the mobile probe should connect
to a real mobile network. Otherwise, an attacker could detect performance
differences (e.g., network delay) in advance. In addition, a connection via
a real mobile operator ensures the assignment of a topological correct IP
address. Note, for an attacker it is easy to identify relevant IP blocks,
either by testing or analysing meta data in the Internet registries.

As we are mainly interested in the analysis of statistical effects and not
on dedicated attacks, the mobile honeypot is primarily based on
low-interaction honeypots.


\subsection{Implementation}

\paragraph{Software}
To implement the proposed honeypot system, we use multiple well-known
honeypot tools. The mobile honeypot consists of the following different
sub-honyepots: Kippo, Glastopf, and Dionaea.

\emph{Kippo} is a dedicated SSH honeypot that emulates remote terminal
sessions. Login access is secured by a trivial password, which allows an
attacker to gain easily access to the system. The user account is granted
administrator privileges. An attacker can execute common programs, as well
as download and install additional tools. The honeypot records downloaded
files in the background for later analysis. To protect the honeypot against
compromising operations, all infiltrated actions are only valid within the current
attack session and the execution of newly installed programs is prohibited.
Note, this does not conflict with our objectives, as we are interested in
the principle behaviour of the attacker.

\emph{Glastopf} implements a web-based media server providing an upload
form. Uploaded data can be stored in a simulated smartphone file system.
This honeypot emulates typical vulnerabilities of a web system.

\emph{Dionaea} is used to emulate TFTP and FTP services.

To detect generic attacks, we use \emph{Honeytrap}. It listens on all other
transport ports and is particularly useful to analyse statistical effects.
Worms, for example, do not need a protocol compliant negotiation of
transmission parameters but send data via an existing TCP connection
without waiting for corresponding replies.


\begin{figure*}
  \subfigure[Mobile network]{\includegraphics[width=0.5\columnwidth]{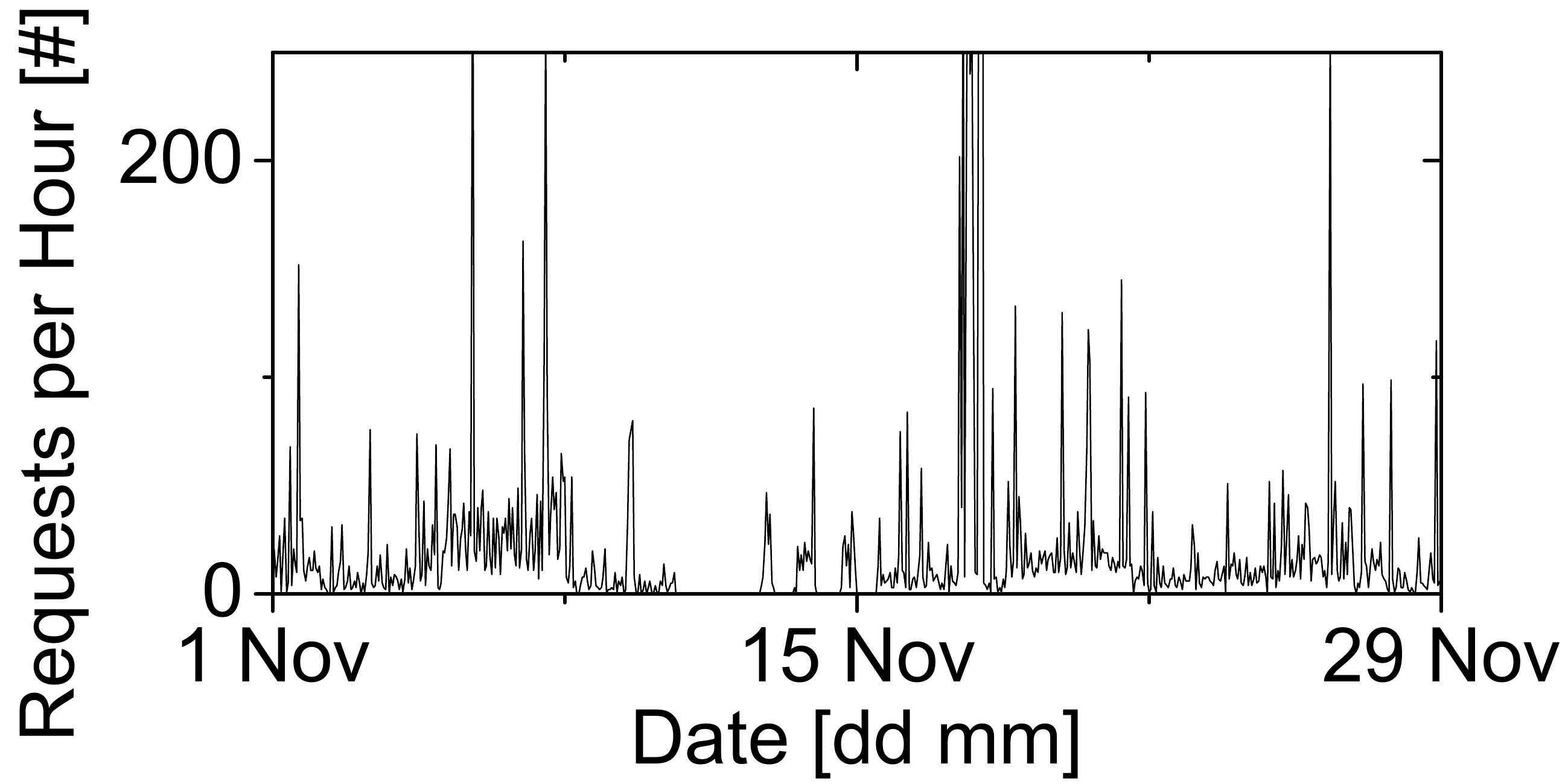}}
  \subfigure[DSL network]{\includegraphics[width=0.5\columnwidth]{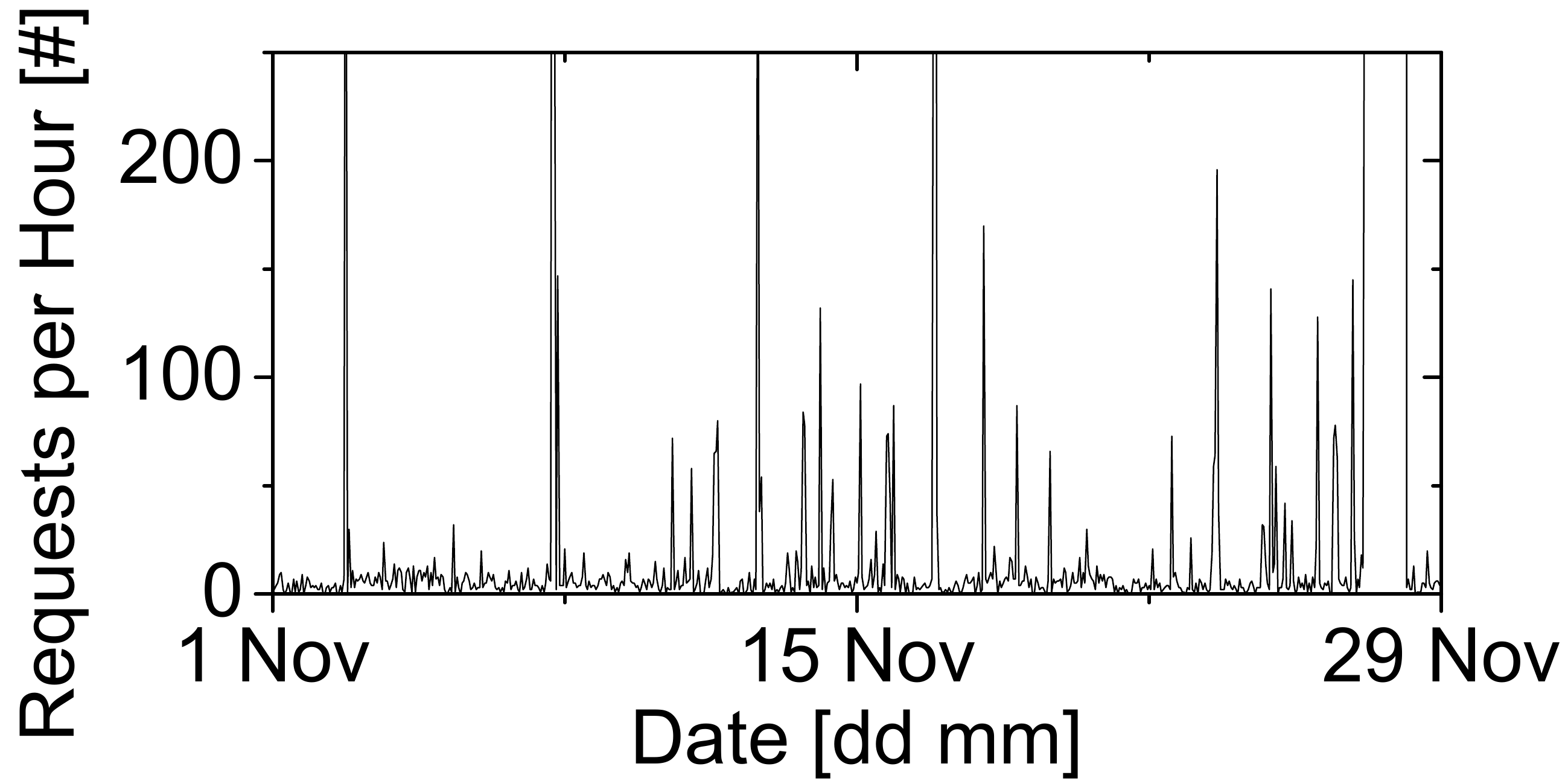}}
  \subfigure[Darknet]{\includegraphics[width=0.5\columnwidth]{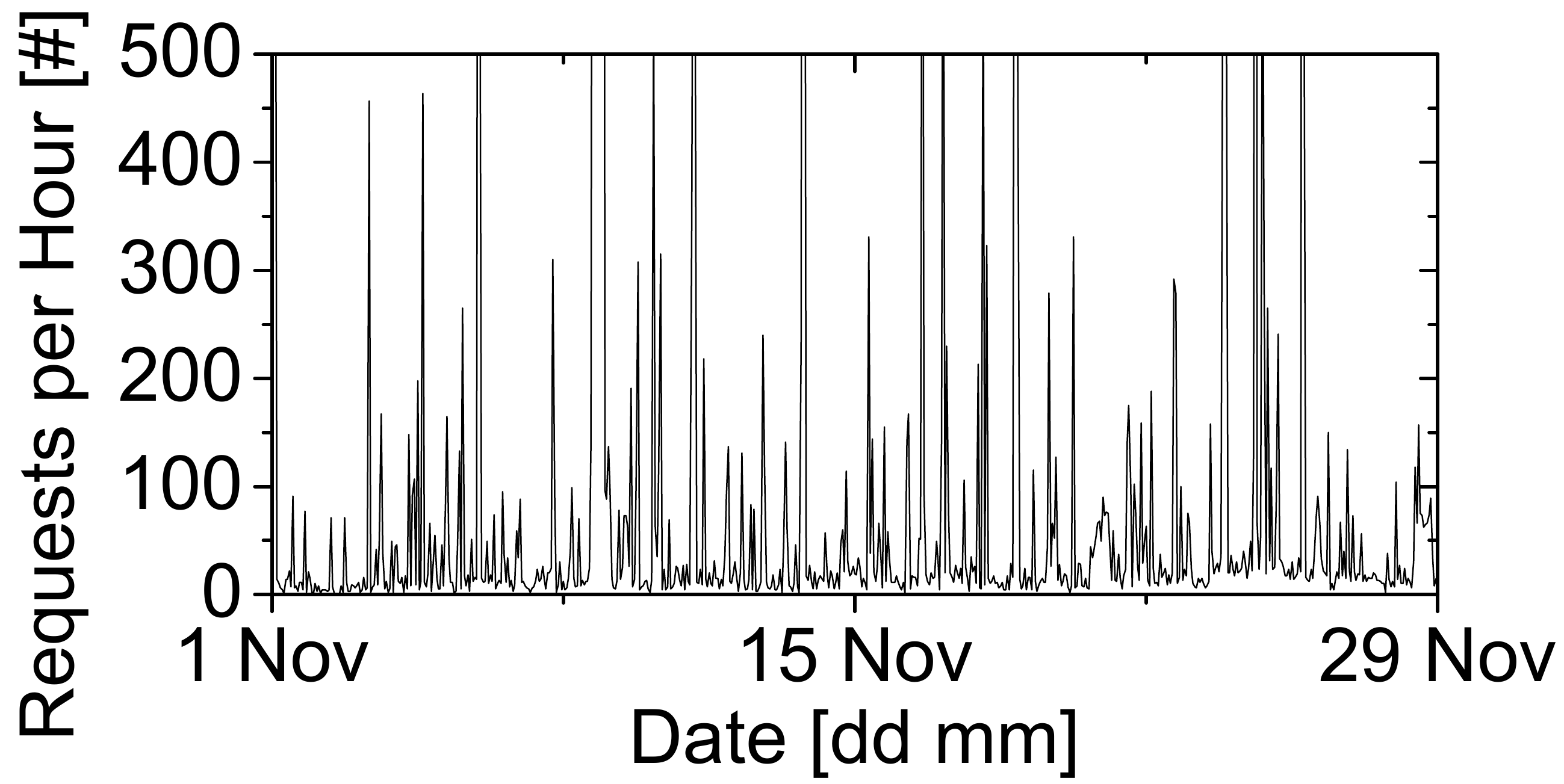}}
  \subfigure[University network]{\includegraphics[width=0.5\columnwidth]{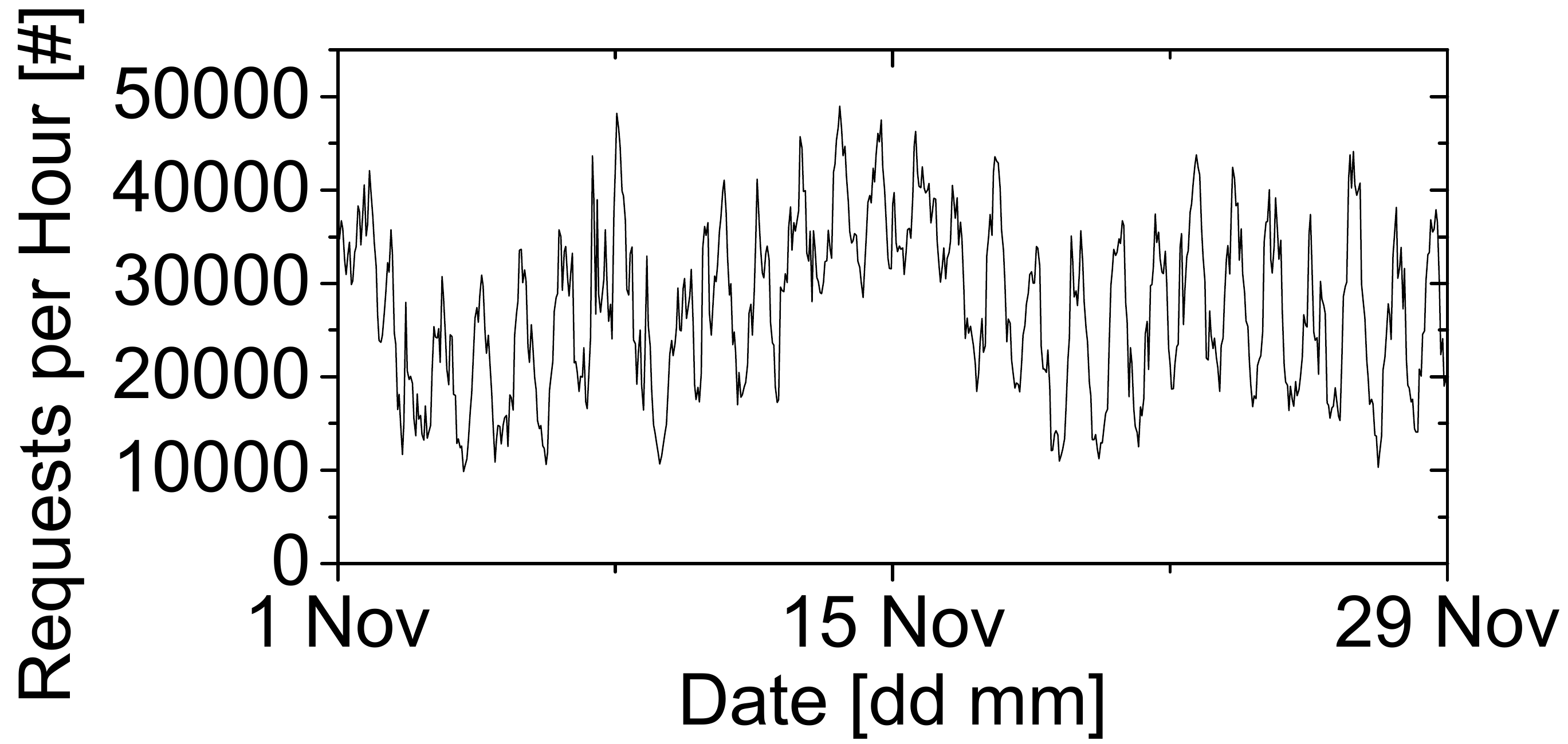}}
  \caption{Comparing amount of attacks on different between mobile and
  non-mobile honeypot probes, Nov. 2012} \label{fig:honeypot-deployment}
\end{figure*}

\paragraph{Network connectivity}
Several mobile operators provide only private IP addresses. Nevertheless,
there is a continuous demand for public IP addresses. In particular with an
increased deployment of IPv6, we expect a significant change, which will
enable mobile nodes to participate in the Internet without NAT traversal.

In addition, many mobile operators, at least in Germany, prevent the
communication between end devices per default in NAT domains. For this
reason, the deployed mobile honeypot presented in this paper is connected
via the Deutsche Telekom, one of the largest telecommunications companies
in Europe. They allow to choose an alternative Access Point Name (APN) that
provides public IP addresses and thus intra-domain communication.

Note, the proposed honeypot system can be connected to any other type of
network access, such as a DSL home network or business Internet access.
This allows us to use the same system in different network environments
(i.e., wired and wireless infrastructures), and to monitor attack behaviour
from different vantage points in the Internet without loosing
reproducibility.



\subsection{Deployment}

We started the deployment of the honeypot systems at common PC hardware mid
of 2011. Since then they run continuously and surprisingly well. The mobile
honeypots of both independent groups include one iOS and two Android
probes. They are connected via an USB stick to the UMTS network. All data
is exported in a five minute interval to the early warning system of one
European's largest telecommunication companies. To prevent interference and
preserve bandwidth of the UMTS link, log data is transmitted using a
separate LAN connection. Data from or to the log server is excluded from
further analysis.

In addition to the measurement probes that use a mobile Internet access,
we deployed the same system at three nodes connected to different
non-mobile networks. In detail, the network access is (1) a university
network, which reflects a stable and well-known open access; (2) a DSL
network, which represents a common home uplink; (3) a darknet, which
highlights background noise, because it does not announce any service.
These characteristic access types allow for comparison of the mobile
measurements with non-mobile environments.

%% file: results.tex
\section{Measurement Study} \label{sec:ergebnisse}

In this section, we present preliminary measurement results of the mobile
honeypot system. We could not find substantial disagreements between the
different mobile probes. We count any external connect to the honeypot
system as an attack, its source IP address is called the attacker.

\begin{table*}
\center
\begin{tabular}{lrrrrrrrr} \toprule
& \multicolumn{4}{c}{\# Attacked ports per transport protocol} & \multicolumn{4}{c}{\# Attacks per transport protocol} \\
\cmidrule(r){2-5} 
\cmidrule(r){6-9}
& UMTS & Darknet & DSL & University & UMTS & Darknet & DSL &
University \\ 
\midrule
TCP & 111 & 133 & 89 & 252 & 14,954 & 55,378 & 32,781 & 22,445,580 \\
UDP & 76 & 71 & 96 & 22 & 637 & 5,583 & 8,254 & 480 \\
\bottomrule
\end{tabular}
\caption{Amount of malicious requests per transport protocol, November
2012} \label{tbl:angriffeport}
\end{table*}

\begin{table*}
\centering
\begin{tabular}{lrrrrrrrrrr}
\toprule
\multicolumn{2}{c}{Rank \hfill 1} & 2 & 3 & 4 & 5 & 6 & 7 & 8 & 9 & 10 \\
\midrule
\multirow{2}{*}{UMTS} & 22 & 1433 & 3306 & 5900 & 6666 & 3389 & \emph{1080} & 23 & \emph{5060} & 80 \\
& SSH & MSSQL & MSSQL & VNC & & RDP & \emph{SOCKS} & Telnet & \emph{SIP} & HTTP \\ \midrule

\multirow{2}{*}{Darknet} & 22 & 139 & 110 & 25 & 3306 & \emph{91} & 5901 & 5900 & 3389 & 53 \\
& SSH & NetBIOS & POP3 & SMTP & MSSQL & & & VNC & RDP & DNS \\ \midrule

\multirow{2}{*}{DSL} & \emph{51099} & 22 & 5900 & 110 & 25 & 3389 &
143 & 6666 & 1433 & 23 \\

& & SSH & VNC & POP3 & SMTP & RDP & IMAP & & MSSQL & Telnet \\ \midrule
\multirow{2}{*}{Univ.} & \emph{445} & 139 & 80 & 22 & 110 & 3389 & 5900 &
3306 & 143 & \emph{5902}
\\
& \emph{MS AD} & NetBIOS & HTTP & SSH & POP3 & RDP & VNC & MSSQL & IMAP & \\
\bottomrule
\end{tabular}
\caption{Top-10 of the most attacked ports (single events emphasized),
November 2012}
\label{tbl:top10ports}
\end{table*}

\subsection{General Observations}

In general, the number of attacks targeting the mobile probe do not
significantly differ from honeypots connected to the Internet via typical
wired access. It seems that the attackers scan the Internet without
considering specific network types but try to exploit as many devices as
possible.

The procedure of the attacker is almost identical to the wired probes.
After gaining successfully a shell login and executing some common commands,
an adversary usually downloads malicious software and tries to integrate
the honeypot into an IRC-botnet. The attacker initiates commands almost
independently of the local system properties even if this leads to conflicts
(e.g., non-existing directories). We frequently observed that the adversary
navigates through the file system directories following the common Linux
structure. Specific Android processes have been ignored.

To our surprise, we observed very rarely an intruder that conducted a
mobile-specific attack. For example, after establishing an SSH connection
to the mobile honeypot, one adversary targeted on the address book as well
as the stored photos of the emulated mobile system. Those attacks are
usually performed manually and not based on scripts. However, the mobile
honeypot did not captured Android- or iOS-specific malware or Exploits.

\subsection{Comparative Detail Analysis}
For our subsequent analysis we focus on network traffic and compare effects
on the mobile honeypot with non-mobile systems. We consider the measurement
period of November 2012. 

Most of the external requests are related to the university network (cf.,
Table~\ref{tbl:angriffeport}). The DSL and UMTS honeypots measure on
average 21 and 55 attacks per hour, respectively. More surprisingly, the
darknet experiences on average about 83 external requests. Around 90\% of
the attacks use TCP. The prominent ports are 22 (SSH), 1433/3306 (MSSQL),
and 80 (HTTP). We summarize details in Table~\ref{tbl:top10ports}.


\subsubsection{Attacks per AS}

To explore the topological location of the attacks, we map the source IP
addresses of the adversaries to their origin autonomous system (AS) using
the common IP to ASN lookup service provided by Team Cymru. We rank each AS
separately per network access.

\begin{figure*}
  \center
  \subfigure[Attacks per AS]{\includegraphics[width=0.9\columnwidth]{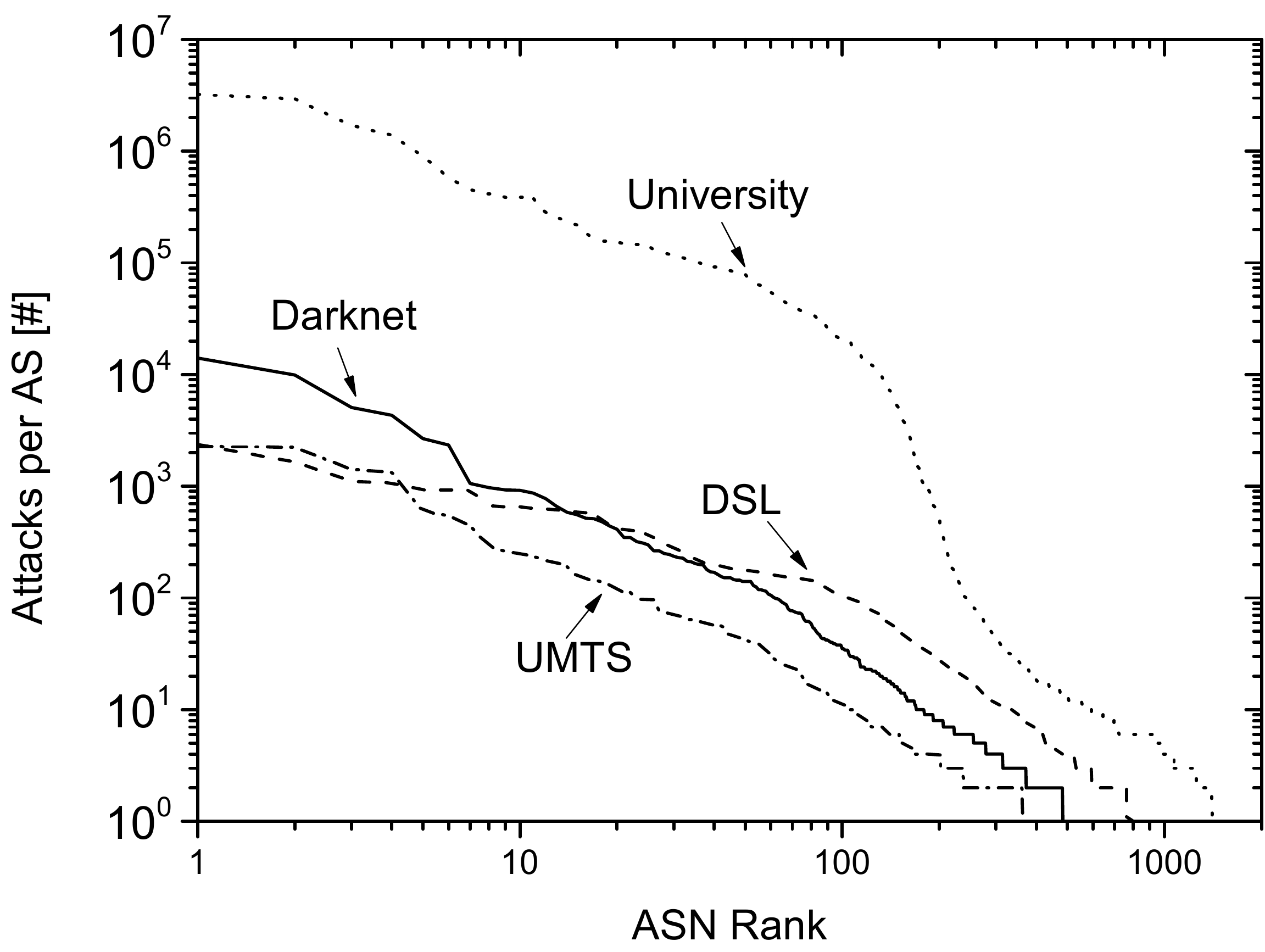}\label{fig:attacksperas}}\quad
  \hfill
  \subfigure[Attackers per AS]{\includegraphics[width=0.9\columnwidth]{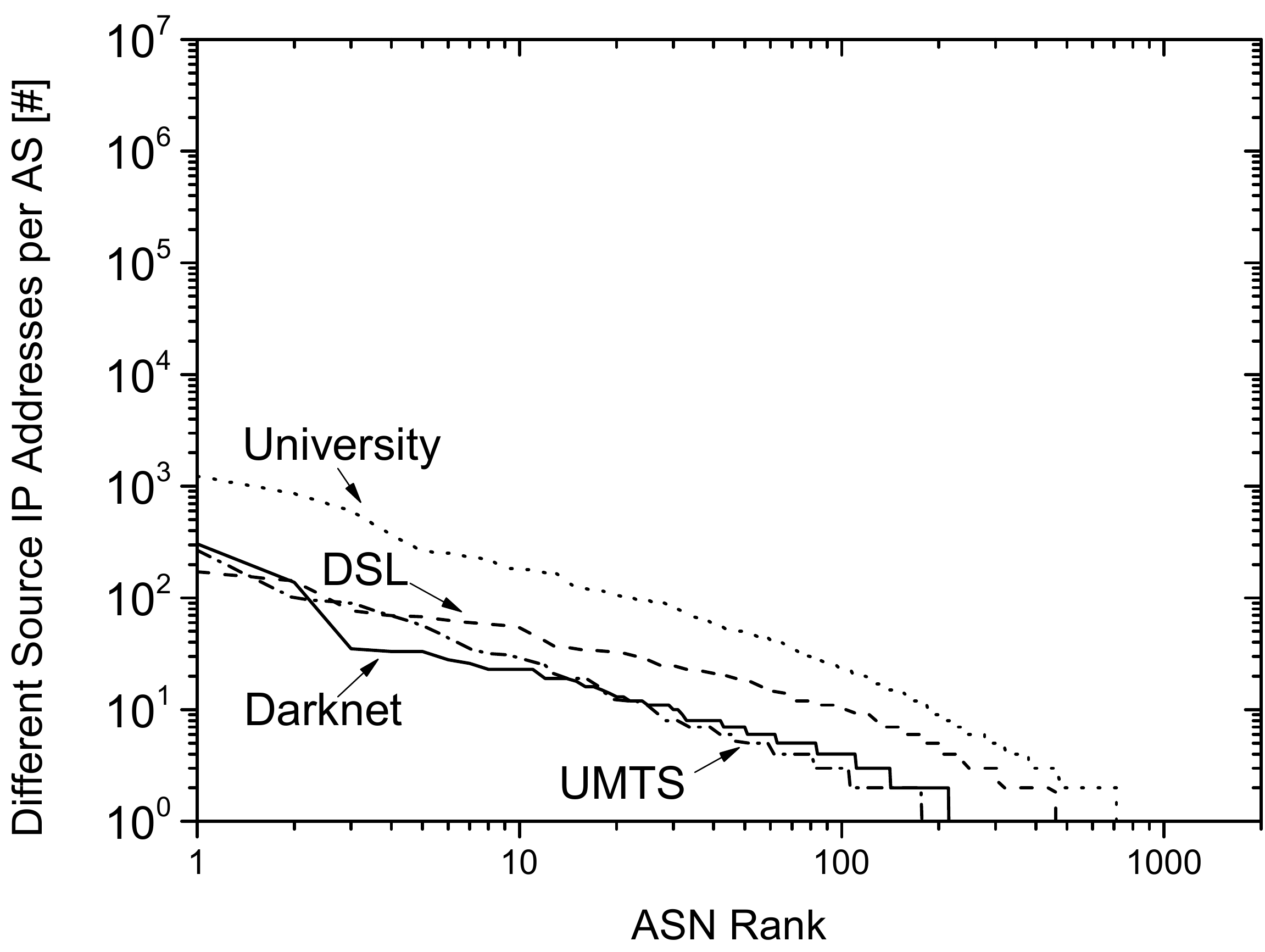}\label{fig:adversariesperas}}
  \caption{Comparison of requests per autonomous system separately ranked
  by network access, Nov. 2012}\label{fig:attackersperas}
\end{figure*}

Overall, most of the attacks are initiated from IP prefixes that belong to
the same small set of ASes (cf., Figure~\ref{fig:attacksperas}). The top-5
ASes are primarily based in China and Russia and do not cover mobile
operators. The distribution of attacks is enhanced for the university
network. The darknet and the DSL home network follow a similar shape, in
which the mobile network exhibits a more narrowed distribution. For all
network types, it is clearly visible that already a small number of ASes
have a significant impact on the attack experiences.

\subsubsection{Attackers per AS}

In our second statistical analysis, we measure the number of different
source IP addresses (i.e., attackers) per AS (cf.,
Figure~\ref{fig:adversariesperas}). Again, we calculate the rank separately
for each network access type.
This analysis allows us to estimate the amount of different attack sources
and to balance the intensity of each attacker. Consequently, the maximal 
values are three to two orders of magnitude less compared to the number of
attacks. Nevertheless, the characteristic shape of the curves in
Figure~\ref{fig:attacksperas} still exists.

\subsubsection{Comparison with Previous Measurements}

In our previous measurements from January 2012 \cite{wtkss-fimh-12} we
found similar results. The most significant difference is the absolute
number of attacks on the mobile system and the darknet. In November 2012
the darknet experienced a surprisingly high amount of attacks, indicating
that it is more attractive to an attacker compared to the UMTS and home
network. This is in general not true as the IP address space of the darknet
is officially not related to any external network service. Looking into
this in more detail reveals that a small set of nodes connected to the
darknet initiating a large portion of requests. This observation is also
highlighted by our analysis per attacker.

Similar to this, the UMTS network is not spared by attacks in general. In
January 2012, the UTMS nodes suffered on average on the same amount of
requests compared to the home network. Interestingly regions and originators
of attacks were better pronounced and operate at higher intensity in the
beginning of this year. We consider this as an indicator for the liveliness
of the mobile regime, which needs further analysis in the~future.


%% file: conclusion.tex
\section{Conclusion and Outlook} \label{sec:zusammenfassung}

In this paper we presented a mobile honeypot system that allows for a
detailed analysis of mobile-specific attacks. A key insight is the
abstraction from unnecessary mobility aspects. To study common attacks, the
honeypot is neither required to run on a real smartphone, nor on a
full-fledged mobile operating system. The mobile honeypot is operated on
standard PCs running Linux. This enables the analysis of malicious traffic
across different network environments and bears the advantage of simplified
long-term maintenance as the same tool basis can be re-used.

We deployed our concept on probes connected to a mobile network, as well as
monitoring nodes connected to different types of wired Internet access
(i.e., university network, darknet, DSL home network). So far we did not
find a relevant ratio of remote attacks that specifically target on the
mobile system, neither from non-mobile nor mobile networks. From this
perspective we conclude that mobile devices are currently more threatened
with malicious applications (e.g., trojan horse) compared to external,
unsolicited requests via the Internet.

In the future we will still maintain our honeypot setup. We will
concentrate on more subtle correlation analysis of how specific groups of
attackers behave with the aim to identify individual patterns of mobility
related aggressions. We will also analyse attacks per port in more detail,
and conduct time series analysis.
Due to limited statistics, though, these considerations will require a much
longer range of observation. Estimating the error of IP spoofing events on
our results is also part of our future work.

%% file: acks.tex
\vspace{-0.15cm}
\paragraph{Acknowledgements}
We would like to thank Marcin Nawrocki for taking care of the mobile
honeypots coordinated by the Freie Universit\"at Berlin. Sebastian Trapp is
gratefully acknowledged for early discussions on this topic.  This work is
partly supported by the German BMBF within the project SKIMS
(http://skims.realmv6.org).